\documentclass[conference]{IEEEtran}
\IEEEoverridecommandlockouts
\usepackage{cite}
\usepackage{amsmath,amssymb,amsfonts}
\usepackage{algorithmic}
\usepackage[dvipdfmx]{graphicx}
\usepackage{bm}
\usepackage{braket}
\usepackage{textcomp}
\usepackage{xcolor}
\def\BibTeX{{\rm B\kern-.05em{\sc i\kern-.025em b}\kern-.08em
    T\kern-.1667em\lower.7ex\hbox{E}\kern-.125emX}}
\usepackage{color}

\begin{document}

\title{Magnetostriction studies up to megagauss fields using fiber Bragg grating technique\\
}

\author{
\IEEEauthorblockN{Akihiko Ikeda}
\IEEEauthorblockA{
\textit{Institute for Solid State Physics,}\\
\textit{University of Tokyo}\\
Kashiwa, Japan \\
ikeda@issp.u-tokyo.ac.jp}
\and
\IEEEauthorblockN{Yasuhiro H. Matsuda}
\IEEEauthorblockA{
\textit{Institute for Solid State Physics,}\\
\textit{University of Tokyo}\\
Kashiwa, Japan}
\and
\IEEEauthorblockN{Daisuke Nakamura}
\IEEEauthorblockA{
\textit{Institute for Solid State Physics,}\\
\textit{University of Tokyo}\\
Kashiwa, Japan}
\and
\IEEEauthorblockN{Shojiro Takeyama}
\IEEEauthorblockA{
\textit{Institute for Solid State Physics,}\\
\textit{University of Tokyo}\\
Kashiwa, Japan}
\and
\IEEEauthorblockN{Hiroshi Tsuda}
\IEEEauthorblockA{
\textit{National Institute of Advanced}\\
\textit{Industrial Science and Technology}\\
Tsukuba, Japan}
\and
\IEEEauthorblockN{Kazuya Nomura}
\IEEEauthorblockA{
\textit{Institute for Solid State Physics,}\\
\textit{University of Tokyo}\\
Kashiwa, Japan}
\and
\IEEEauthorblockN{Ayumi Shimizu}
\IEEEauthorblockA{\textit{Institute for Solid State Physics,}\\
\textit{University of Tokyo}\\
Kashiwa, Japan}
\and
\IEEEauthorblockN{Akira Matsuo}
\IEEEauthorblockA{
\textit{Institute for Solid State Physics,}\\
\textit{University of Tokyo}\\
Kashiwa, Japan}
\and
\IEEEauthorblockN{Toshihiro Nomura}
\IEEEauthorblockA{\textit{Hochfeld-Magnetolabor Dresden,}\\
\textit{Helmholtz-Zentrum Dresden-Rossendorf}\\
Dresden, Germany}
\and
\IEEEauthorblockN{Tatsuo C. Kobayashi}
\IEEEauthorblockA{\textit{Department of Physics}\\
\textit{Okayama University}\\
Okayama, Japan}
\and
\IEEEauthorblockN{Takeshi Yajima}
\IEEEauthorblockA{
\textit{Institute for Solid State Physics,}\\
\textit{University of Tokyo}\\
Kashiwa, Japan}
\and
\IEEEauthorblockN{Hajime Ishikawa}
\IEEEauthorblockA{
\textit{Institute for Functional Matter}\\
\textit{and Quantum Technologies}\\
\textit{University of Stuttgart}\\
\textit{Stuttgart,Germany}}
\and
\IEEEauthorblockN{Zenji Hiroi}
\IEEEauthorblockA{
\textit{Institute for Solid State Physics,}\\
\textit{University of Tokyo}\\
Kashiwa, Japan}
\and
\IEEEauthorblockN{Masahiko Isobe}
\IEEEauthorblockA{\textit{Max Planck Institute for}\\
\textit{Solid State Research}\\
Stuttgart, Germany}
\and
\IEEEauthorblockN{Touru Yamauchi}
\IEEEauthorblockA{
\textit{Institute for Solid StaUnivte Physics,}\\
\textit{University of Tokyo}\\
Kashiwa, Japan}
\and
\IEEEauthorblockN{Keisuke Sato}
\IEEEauthorblockA{
\textit{National institute of Technology,}\\
\textit{Ibaraki College}\\
Ibaraki, Japan}
}

\maketitle

\begin{abstract}
We here report magnetostriction measurements under pulsed megagauss fields using a high-speed 100 MHz strain monitoring system devised using fiber Bragg grating (FBG) technique with optical filter method.
The optical filter method is a detection scheme of the strain of FBG, where the changing Bragg wavelength of the FBG reflection is converted to the intensity of reflected light to enable the 100 MHz measurement.
In order to show the usefulness and reliability of the method, we report the measurements for solid oxygen, spin-controlled crystal, and volborthite, a deformed Kagom\'{e} quantum spin lattice, using static magnetic fields up to 7 T and non-destructive millisecond pulse magnets up to 50 T.
Then, we show the application of the method for the magnetostriction measurements of CaV$_{4}$O$_{9}$, a two-dimensional antiferromagnet with spin-halves, and LaCoO$_{3}$, an anomalous spin-crossover oxide, in the megagauss fields.

\end{abstract}


\begin{figure*}[t!]
\centerline{\includegraphics[width=18cm]{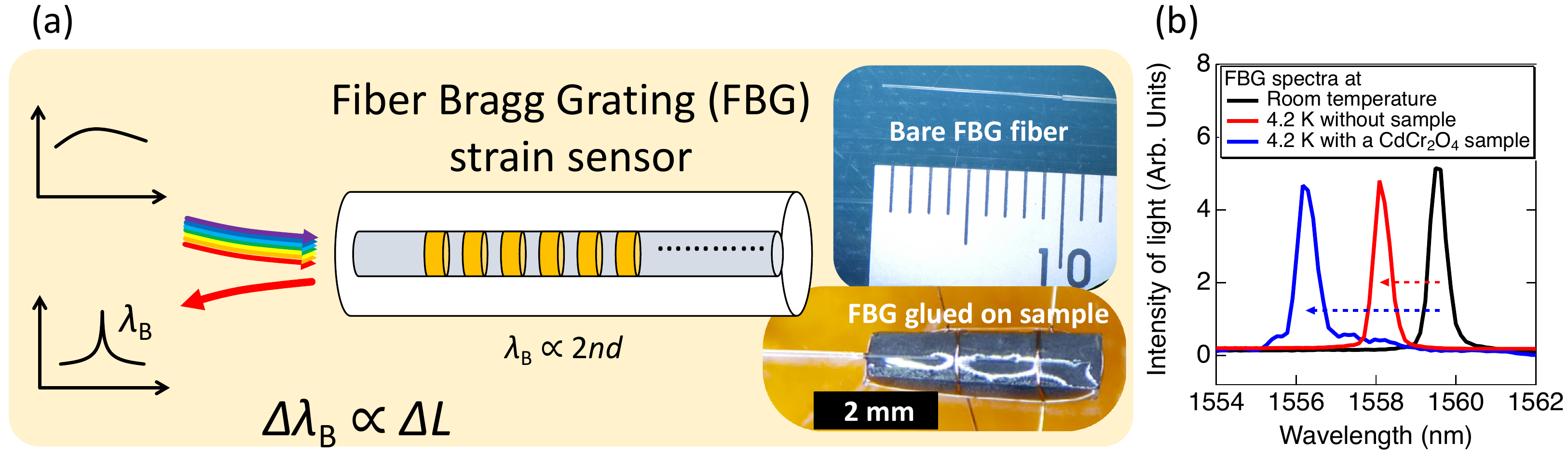}}
\caption{(a) Schematic drawing explaining how a fiber Bragg grating (FBG) strain sensor works with pictures of bare FBG and FBG glued to a sample.
(b) Spectra of FBG reflection at room temperature, at 4.2 K without and with a CdCr$_{2}$O$_{4}$ sample, indicating that the FBG follows the strain of the sample.}
\label{fbg}
\end{figure*}

\section{Introduction}
Magnetic field is a useful tool to investigate the nature of matter by directly acting on spin, orbital and phases of electrons.
Ultrahigh magnetic fields reaching 100-1000 T (=1-10 Megagauss) have very large Zeeman energy that well exceeds that of the room temperature \footnote{1000 T corresponds to 1343 K with spin-1/2 and $g=2$}.
Such high magnetic field is used not only for better characterization of matter in resonance methods but also used for bringing about emergent quantum phase of matter that is stable  even at room temperature.
Besides, a number of non-magnetic materials can be made into magnetic ones accompanying drastic change of physical properties (conductivity, magnetization and optical responses) and even their chemical properties.

Recently, magnetic fields reaching or approaching 100 T with millisecond pulse duration have been generated in non-destructive manner in several facilities in the world, being used for solid state researches.
On the other hand, megagauss fields with much shorter pulse duration (a few micron seconds) have been generated in destructive manner using flux compression methods and fast discharge of capacitor bank for decades \cite{herlach1968, fowler, herlach}.
Measurements of condensed matter properties in such destructively generated megagauss fields severely suffer from the limited time and space of the generated magnetic fields. Due to the nature of the megagauss pulse magnets such as electromagnetic flux compression (EMFC) technique and single turn coil (STC) method \cite{miura}, measurement techniques are required to be robust against electric noises and to respond in high-speed at $>$10 MHz under the $\mu$s-pulsed megagauss field pulse.
Numerous efforts have been paid for the development of measurement techniques in megagauss fields such as conductivity measurements using radio frequency (RF) transmission \cite{SekitaniRF, SuyonRF} and RF resonance \cite{NakamuraRF} techniques, magnetization measurements using faraday rotation \cite{Nojiri, MiyataPRL} and induction methods \cite{Amaya, TakeyamaM01, TakeyamaM02, IkedaPRB01}, and so forth.

Number of novel findings are made recently in megagauss regions in condensed matter physics.
One of common concerns among them is the spin-lattice coupling.
For example, a new phase is found for solid oxygen at megagauss region \cite{Nomura01}. 
Oxygen is a molecular magnet with spin-one.
Upon cooling, oxygen gas condense to liquid and then to several solid phases, whose lattice and magnetism are strongly coupled with each other thanks to the comparable energy scale of magnetic exchange interaction and inter-molecular Van der Waals interaction.
At the lowest temperature, $\alpha$ phase has an antiferromagnetic long-range order.
Application of megagauss fields collapses the antiferromagnetism and brings about a new high-field phase, $\theta$ phase \cite{Nomura01}.
The $\theta$ phase is identified by magnetization measurement and optical spectroscopy.
It is strongly anticipated that the new $\theta$ phase accompanies the drastic change of the lattice symmetry.
However, no direct experimental observation of the change of the lattice is reported so far.

Geometrically frustrated spin system on pyrochlore lattice of CdCr$_{2}$O$_{4}$ \cite{MiyataPRB} and  ZnCr$_{2}$O$_{4}$ \cite{MiyataPRL} are investigated in megagauss region, where the spin-lattice coupling plays a crucial role.
Kinds of magnetic phases are found in its magnetization process from the ordered ground state to the full saturation.
Anomaly just before the saturation is argued to be related to the spin nematic order, which is a long range order of quadruple moment of spin .
Such exotic phase is argued to be stabilized by the spin-lattice coupling via the bi-quadratic term in the spin Hamiltonian.
Lattice effects in the spin nematic phase is non-trivial and never been investigated in detail, with an exception of a x-ray diffraction study \cite{Tanaka}.

Low dimensional quantum spin system exhibits quantum phases that is absent in classical spin systems.
SrCu$_{2}$(BO$_{3}$)$_{2}$ is a two-dimensional quantum spin system on a Shastry-Sutherland lattice.
The ground state is a gapped nonmagnetic phase, which arise from the formation of singlet spin dimers.
The spin gap is collapsed by application of high magnetic fields, leading to formation of number of ordered phases. 
By magnetostriction measurements, it was found that the longitudinal magnetostriction curve in $c$ axis is proportional to the magnetization curve \cite{JaimePNAS, Radtke}.
This indicates that even in the quantum magnets, where the spin-lattice couplings are small, one can explore magnetic phases though magnetostriction measurements.
This is helpful because, at megagauss region, magnetostriction measurement can be a good complement for the magnetization measurement.
Another valuable point of magnetostriction measurements of quantum magnets is that strain is related to spin correlation as $\epsilon = - c\braket{\hat{\bm{S}}_{i}\cdot\hat{\bm{S}}_{j}}$, where $c$ is a constant related to exchange constant and elastic constant.
The spin correlation includes not only the $S^{z}$ but also $S^{x}$ and $S^{y}$, which are difficult to be observed  with other techniques.
The fact makes magnetostriction sensitive to Bose-Einstein condensation of magnons where the perpendicular components of  spin are fixed in a direction throughout the material.
There is one report on the observation of the order parameter of magnon BEC with magnetostriction measurement \cite{vivian}.

\begin{figure*}[t!]
\centerline{\includegraphics[width=18cm]{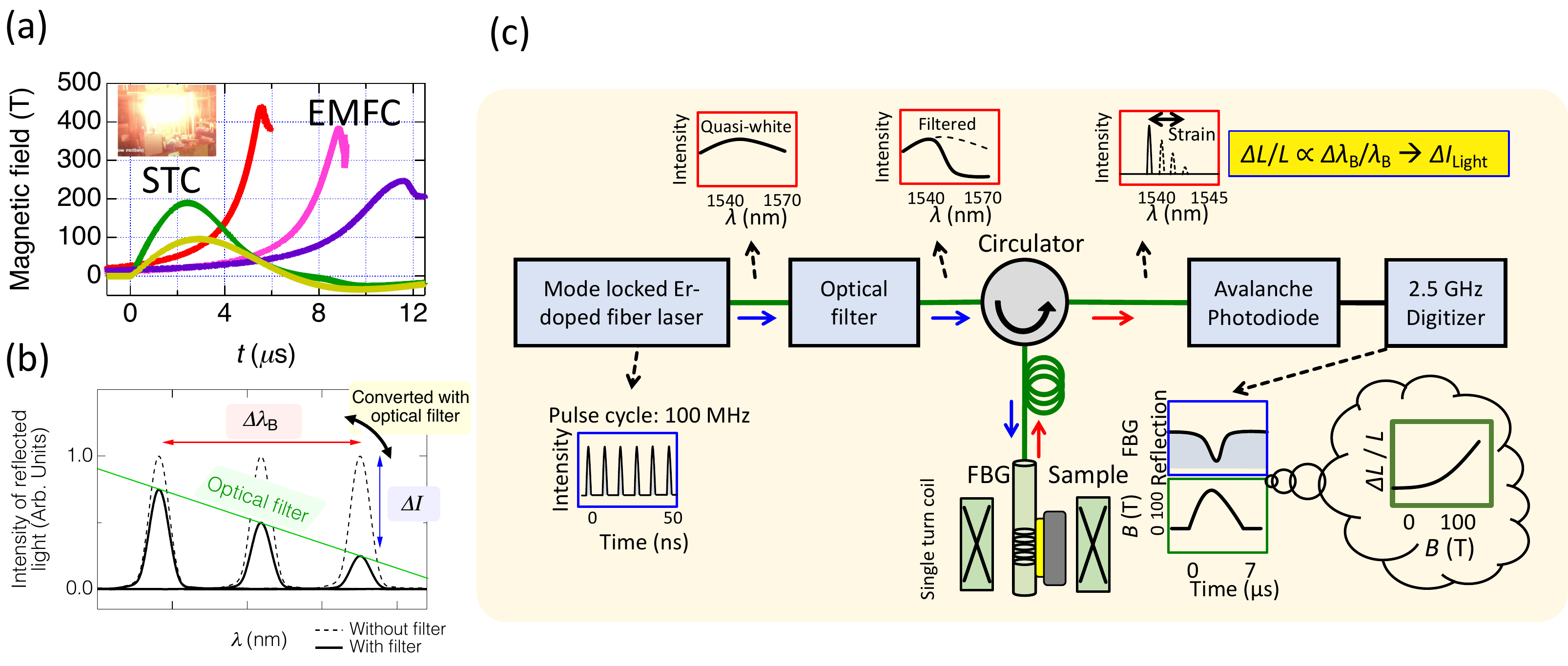}}
\caption{(a) Typical waveform of megagauss pulsed magnetic field generated using destructive pulse magnets, single turn coil (STC) and electromagnetic flux compression (EMFC) methods.
(b) Schematic drawing explaining how the optical filter method works as a detection scheme of the FBG based strain monitoring.
The shift of the Bragg wavelength $\Delta\lambda_{\rm{B}}$ is converted to the intensity of reflected light $\Delta I$ using an optical filter.
(c) An actual measurement setup of the FBG based high-speed 100 MHz strain monitoring system using the optical filter method.
The intensity of light reflected from the FBG is monitored using an avalanche photo-diode and recorded using an 2.5 GHz oscilloscope along with the pulsed magnetic field.
}
\label{setup}
\end{figure*}

Electronic states can be altered by applications of high magnetic fields in strongly correlated electron systems such as spin crossover cobaltites, heavy fermion systems and valence fluctuating intermetallics,  leading to a large lattice volume changes.
In a spin crossover cobaltite, LaCoO$_{3}$, the spin-state of the trivalent cobalt ion is classified into low, intermediate and high spin states according to the total spin $S=0$, 1, 2, respectively.
Many electron configurations of these spin states are quite different in the electron occupation number of the $e_{g}$ orbitals, resulting in the large difference in their ionic radius.
We have found magnetic field induced spin-crossover in cobaltites in megagauss regions \cite{IkedaPRB01, IkedaPRB02}, which accompanies very large lattice changes \cite{IkedaRSI2017}.

As for heavy fermion system, the $c$-$f$ hybridization between the localized $f$ electrons and the conduction electrons promotes the formation of Kondo lattice at low temperatures.
The $c$-$f$ hybridization is effective electron transfer from the $f$ orbital to the conduction band.
The valence of $f$ orbital indicates the strength of the Kondo coupling, which is coupled to the ionic radius of $f$ orbitals.
Valence fluctuation is the state where the $c$-$f$ hybridization is so strong that a major portion of $f$ electron is transformed to the conduction electrons. 
When a valence fluctuating state is altered by high magnetic fields, a large lattice volume change is expected to take place accompanying the change of the $f$ valence.
Some system have a gap called Kondo gap with spin and charge gap.
For a Kondo insulator YbB$_{12}$, high magnetic field of 50 T suppress the insulator phase leading to the simultaneous meta-magnetic and metal-insulator transition, where strong Kondo coupling is still present \cite{TerashimaPRL}.
Further application of megagauss magnetic fields induces another magnetic transition at above 100 T \cite{TerashimaJPSJ}.
This transition is considered to be the complete breakdown of the Kondo state and large change of valence at Yb is expected.

As discussed above, the observation of lattice degree of freedom at megagauss fields are quite essential to investigate variety of physical phenomena.
From the technical point of view, one promising technique for the strain measurements at megagauss fields is fiber Bragg grating (FBG) strain sensor,  as is schematically drawn in Fig. \ref{fbg}(a).
Though capacitance bridge method and resistive strain gauge are used with high sensitivity and reliability under static and pulsed magnetic fields below 100 T, it is difficult to be used under megagauss fields because of the harsh electric noises, limited sample space and the needs for calibrations.
FBG is an optical sensor that detect a macroscopic strain of matter.
FBG is immune to noises.
High-speed responses are expected.
FBG is easily glued onto the sample.
The strain of the sample is transmitted to the strain of FBG as shown in Fig. \ref{fbg}(b) down to a low temperature of 4.2 K.
FBG is successfully employed in the non-destructive pulse magnets up to 100.75 T \cite{Daou, JaimePNAS, Rotter, JaimeUO2},  whose measurement speed is up to 50 kHz.

For measurements under megagauss fields, repetition rate exceeding 10 MHz is required due to the short pulse width as mentioned earlier.
A technique is reported for 100 MHz detection of FBG strain monitor that employs a sophisticated technique where optical spectrum is converted to the time domain.
The time spectrum is detected using a high-speed digitizer with 25 GHz bandwidth \cite{Rodriguez}.
We have developed another methodology of 100 MHz detection of FBG strain monitor where we utilize optical filter method.
The high-speed 100 MHz strain sensor based on FBG and optical filter method is quite simple and useful in magnetostriction measurements under millisecond pulsed magnetic fields \cite{IkedaRSI2018} and micron-second megagauss fields \cite{IkedaRSI2017, IkedaPhysica}.

\begin{figure*}[t!]
\centerline{\includegraphics[width=16cm]{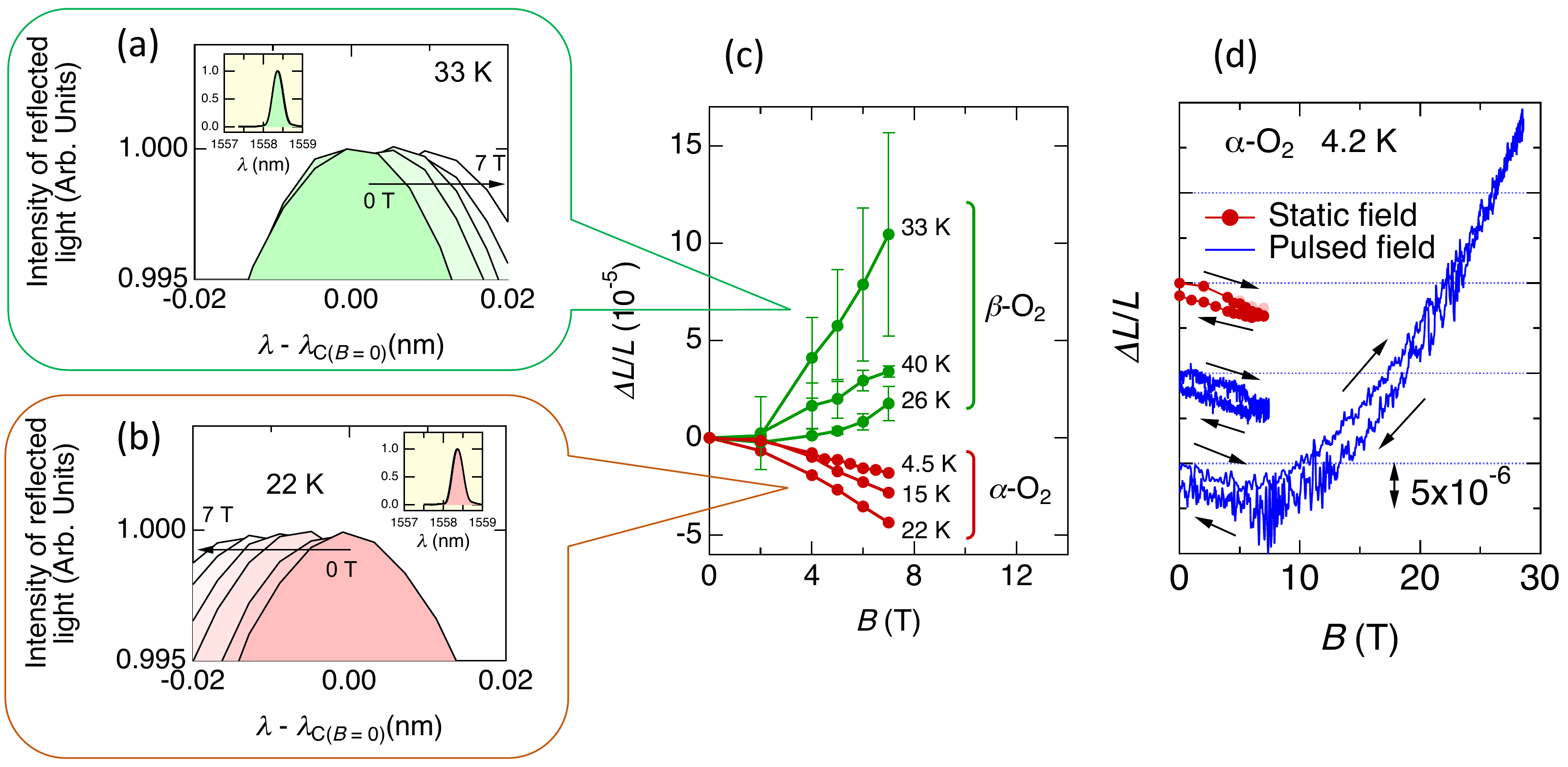}}
\caption{Magnetic field dependence of the reflected spectra of FBG immersed in (a) $\beta$-O$_{2}$ and (b) $\alpha$-O$_{2}$. (c) Magnetostriction of $\alpha$-O$_{2}$ and $\beta$-O$_{2}$ deduced from the $\Delta \lambda_{\rm{B}}/\lambda_{\rm{B}}$ in (a) and (b). (d) Magnetostriction of $\alpha$-O$_{2}$ up to 7 T and 30 T measured using FBG with optical filter method, former of which is compared to the data deduced from the direct observation of FBG spectra.}
\label{oxygen}
\end{figure*}

Here we introduce a high-speed 100 MHz FBG based strain monitor using optical filter method for the detection scheme for the measurement of magnetostriction at megagauss regions.
The method enables us to measure magnetostriction at 100 MHz with a high resolution of $2\times10^{-5}$ at cryogenic temperature down to 2 K.
We show practical applications of the method for solid oxygen, volborthite, CaV$_{4}$O$_{9}$ and LaCoO$_{3}$.


\section{Method}
Typical waveform of the magnetic field generated using destructive pulse magnets of STC and EMFC in ISSP, UTokyo are shown in Fig. \ref{setup}(a).
Recently, we have succeeded in generating an indoor record high field of 12 MG using an EMFC with newly installed capacitor bank system \cite{12MG}.

For magnetostriction measurements under megagauss fields, we utilized the optical filter method as a detection scheme of strain through the reflection from FBG for the high speed measurement.
Detailed description of the measurement system is given in Refs. \cite{IkedaRSI2017, IkedaRSI2018, IkedaPhysica}.
Here we briefly introduce the setup of the optical filter method.
As schematically shown in Fig. \ref{setup}(b), the information of the shift of the Bragg wavelength $\Delta \lambda_{\rm{B}}$ of FBG is converted to the intensity of light by using an optical filter.
Actual setup is shown in Fig. \ref{setup}(c) where we used a mode-locked Er-doped fiber laser as a bright broadband light source.
The transient intensity of light reflected by the FBG is monitored using an avalanche photodiode along with the magnetic field.

The ultrahigh magnetic fields were generated using a horizontal and a vertical STCs in ISSP, UTokyo \cite{miura}. The cryostats used were He-flow \cite{Amaya} and He-bath types \cite{TakeyamaM01} for the horizontal and the vertical STC, respectively.

\section{Results}

\begin{figure*}[t]
\centerline{\includegraphics[width=18cm]{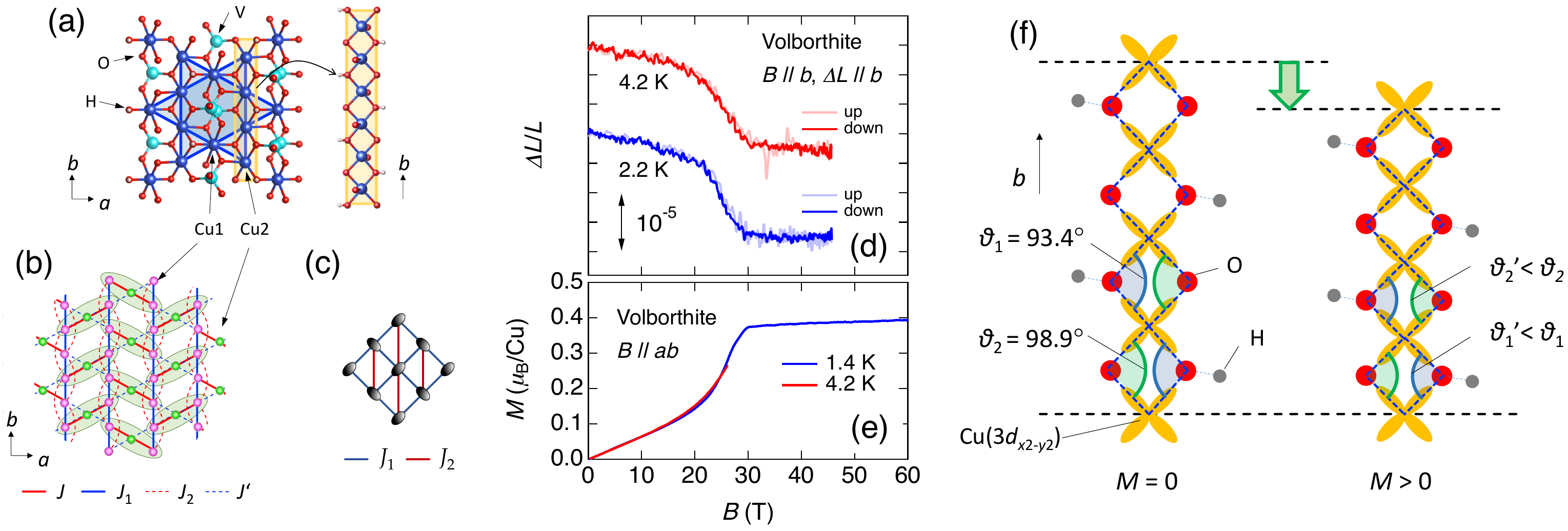}}
\caption{(a) Lattice model of volborthite (Left) and an extracted CuOCu chain (Right) which is responsible for the ferromagnetic exchange constant.
(b) Drawing for the spin trimer model of volborthie \cite{JansonPRL}.
(c) Effective spin model of volborthite deduced from the spin trimer model, where a spin trimer is reduced to a lattice site with spin-half.
The effective model is a two dimensional extension of the one dimensional $J_{1}$-$J_{2}$ model.
(d) Magnetostriction of volborthite measured in the preset study using FBG with optical filter method.
(e) Magnetization of volborthite \cite{YoshidaPRB}.
(f) Schematic model for the model of pantograph like lattice change in the Cu-O-Cu chain with the exchange striction model.}
\label{volborthite}
\end{figure*}

\subsection{Solid oxygen: a spin controlled crystal}

Solid oxygen is a typical spin-lattice coupled system where the crystal structure is determined by the competition between intermolecular forces arising from Van der Waals interaction, electric quadrupole-quadrupole interactions and magnetic exchange interaction between spin angular momentum of O$_{2}$ with spin-one \cite{Freiman2004, Freiman2018}.
At ambient pressure, oxygen undergoes successive phase transitions at 61, 55, 44, 24 K, which correspond to the gas-liquid, liquid-$\gamma$, $\gamma$-$\beta$ and $\beta$-$\alpha$ phase transitions.
All these phase transition accompany simultaneous change of lattice symmetry and magnetic susceptibility indicating strong spin-lattice coupling playing major role in these phase transitions.
Magnetic exchange interactions are antiferromagnetic dynamical exchange interactions between spin-ones on oxygen molecules.
Eventually, the magnetic system falls into antiferromagnetic long-range ordered phase in the $\alpha$-O$_{2}$, where the O$_{2}$ molecules align with collinear molecular orientation perpendicular to its $ab$ plane so that the exchange interaction is maximized by the largest overlap of $\pi$-orbitals.
Whereas, in the $\beta$-O$_{2}$, the antiferromagnetic long-range order is suppressed into a short-range one due to the geometrical frustration effect with the formation of triangular lattice in $ab$ plane, which is deformed in the $\alpha$-O$_{2}$.
$\gamma$-O$_{2}$ is a high-entropy phase in the senses that they are paramagnetic and that the molecular axis of O$_{2}$ is not completely fixed, being called a plastic phase. 

Recently, it was found that the application of megagauss fields on $\alpha$-O$_{2}$ and $\beta$-O$_{2}$ can destroy the antiferromagnetism inducing phase transitions to a novel $\theta$-O$_{2}$, which is experimentally evidenced using magnetization and optical spectroscopy \cite{Nomura01, Nomura02, Nomura03, Nomura04}.
It is claimed that the emerged $\theta$-O$_{2}$ presumably accompanies an alternative lattice symmetry from those of $\alpha$-O$_{2}$ and $\beta$-O$_{2}$.
It is discussed that the antiferromagnetic exchange interaction can be significantly suppressed by changing the molecular orientation from a collinear geometry to the crossed type geometry, which leads to a ferromagnetic exchange interaction being favorable in the momentfull $\theta$-O$_{2}$.
Moreover, the cross type orientation is realized in the ground state of solid nitrogen which is non-magnetic and results from the competing intermolecular Van der Waals interaction and electric quadrupole-quadrupole interactions.
So far, there has been no direct evidence that the phase transitions to the $\theta$-O$_{2}$ accompany a structural change.

As a preceding step for observing structural change in $\alpha$-$\theta$ phase transition, we have carried out a series of magnetostriction measurements using FBG technique for $\alpha$-O$_{2}$ and $\beta$-O$_{2}$.
The results up to 7 T using direct measurement of FBG spectra is shown in Figs. \ref{oxygen}(a), \ref{oxygen}(b) for $\beta$-O$_{2}$ and $\alpha$-O$_{2}$, respectively.
$\Delta L/L$ is deduced from the directly observed $\Delta \lambda_{\rm{B}}/\lambda_{\rm{B}}$ in the FBG spectra and is summarized in Fig. \ref{oxygen}(c).
The striking feature here is the opposite sign of magnetostriction in $\alpha$-O$_{2}$ and that in$\beta$-O$_{2}$, namely, $\alpha$-O$_{2}$ show negative magnetostriction whereas $\beta$-O$_{2}$ show positive ones.

We further measured the magnetostriction of $\alpha$-O$_{2}$ up to 30 T using a non-destructive pulse magnet with millisecond pulse duration.
We employed here the optical filter method as a detection scheme.
In Fig. \ref{oxygen}(d), the result up to 7 T using optical filter method with the pulse magnet is compared with the results using the direct observation of the FBG spectra, which seems to show a very good agreement with each other indicating the validity of the optical filter method as an alternative method of the observation of the strain with FBG.
Beyond 7 T, $\alpha$-O$_{2}$ shows an up turn of the magnetostriction.
The overall magnetostriction curve is analyzed with a form, 
\begin{equation}
\Delta L/L = c_{1} B + c_{2} B^{2}
\end{equation}
with conditions of $c_{1}<0$ and $c_{2}>0$.

We argue that the origin of the quadratic term $c_{2} B^{2}$ is the exchange striction, where increasing magnetization requires the antiferromagnetic exchange interactions to be suppressed inducing the increment of the inter-molecular separation.
On the other hand, the origin of the linear term $c_{1} B$ with the negative coefficient is not clear at this moment.

A possibility is that the linear magnetostriction, which is prohibited due to the time reversal symmetry for paramagnets, may appear as a converse of the piezomagnetism
The presence of the magnetic order with the 66 magnetic point groups that do not contain the symmetry operations of time reversal ($\bar{\bm{1}}$) or time reversal times inversion ($\bar{\bm{1}}'$) can exhibit cross correlated phenomena of strain and magnetization at proximity of zero magnetic field \cite{Borovik, JaimeUO2}.
Detailed discussion will be presented elsewhere.

\begin{figure*}[t]
\centerline{\includegraphics[width=16cm]{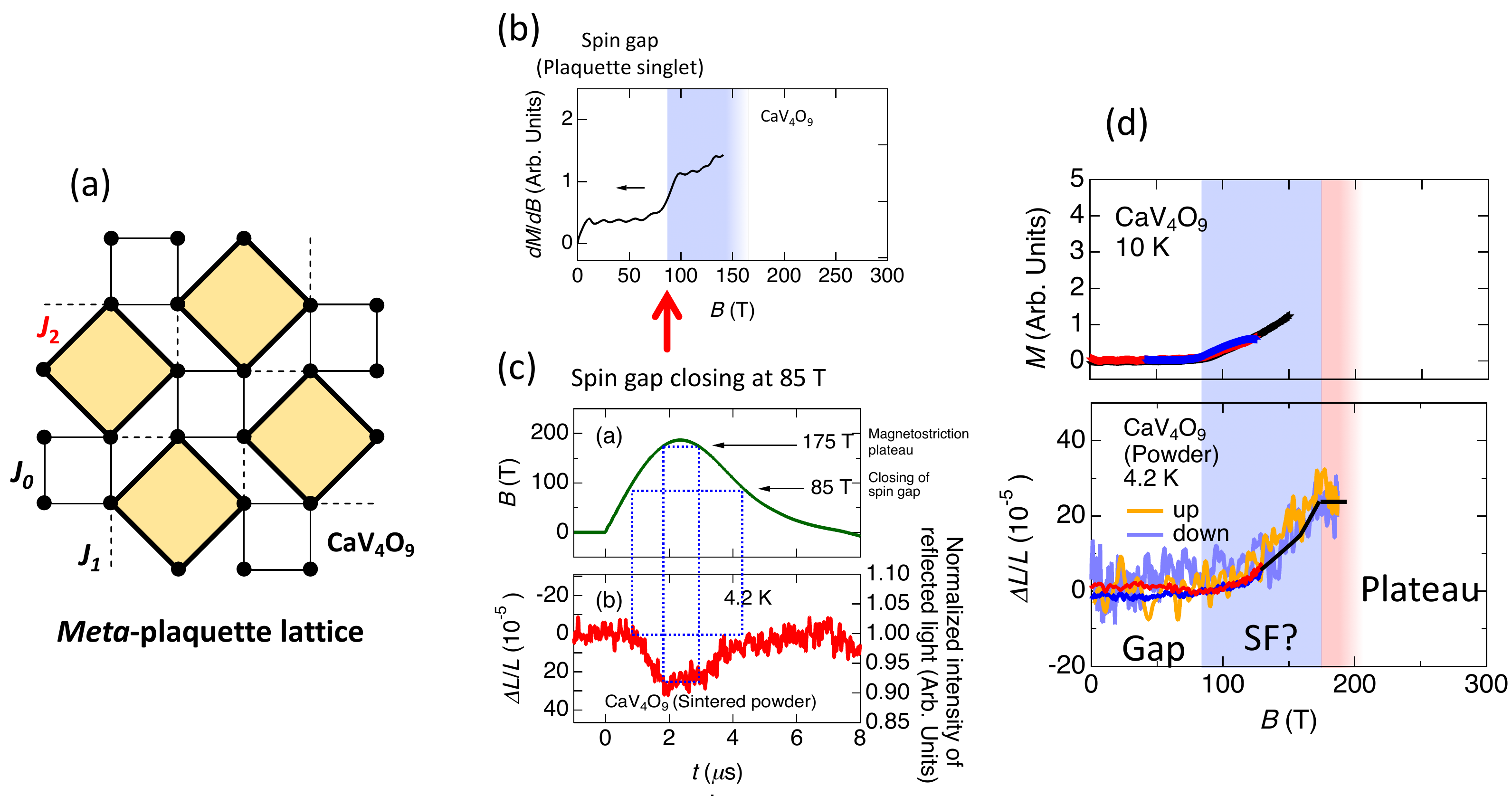}}
\caption{(a) Effective magnetic model of CaV$_{4}$O$_{9}$ with a \textit{meta}-plaquette lattice.
(b) Measured $dM$/$dB$ data of CaV$_{4}$O$_{9}$ up to 150 T showing the collapse of the spin gap at 85 T.
(c) Measured magnetostriction of CaV$_{4}$O$_{9}$ up to 190 T showing the collapse of the spin gap at 85 T and the plateau region above 175 T.
(d) Magnetization and magnetostoriction curves of CaV$_{4}$O$_{9}$.
}
\label{cavo}
\end{figure*}

\subsection{Volborthite: antiferromagnetic quantum spin system on a deformed Kagom\'{e} lattice}
Physics of low dimensional quantum spin system is further enriched by the geometrical frustration that is inherent in triangular, Kagom\'{e} and Maple leaf lattices.
Volborthite was first considered as a candidate for a Kagom\'{e} quantum spin lattice.
Now it is believed that, in volborthite, the deformation of the Kagom\'{e} lattice is significant that the spin network is completely different from the Kagom\'{e} antiferromagnet but is effectively closer to the spin system with nearest neighbor exchange of $J_{1}>0$ and the next nearest neighbor antiferromagnetic exchange constant of $J_{2}<0$ \cite{JansonPRL}.
Thanks to the nearest neighbor ferromagnetic exchange interaction, the spin nematic phase  is claimed to be present at high magnetic fields at around 23-28 T at low temperatures below 2 K.

We have carried out magnetostriction measurements of volborthite up to 45 T at 4.2 K and 2.2 K as shown in Fig. \ref{volborthite}(d).
They are compared with the magnetization curves as shown in \ref{volborthite}(e).
Overall magnetic field dependence of the magnestostriction is in good agreement with the magnetization curve.
A feature is that the negative magnetostriction is observed in the $b$ axis.
Another feature to note is that the magnetostriction curve is best fitted to the magnetization curve to the power of 1.3 as $\Delta L\propto M^{1.3}$.

As a possible origin of the negative magnetostriction, we propose a pantograph like lattice change in the $b$ axis, in which direction the Cu-O-Cu chain extends as shown in Fig. \ref{volborthite}(a) and \ref{volborthite}(f).
In the model, the lattice shows shrinkage in $b$ direction with increasing magnetization, which is driven by the reduction of the Cu-O-Cu angle approaching 90$^{\circ}$ from the initial value greater than $90^{\circ}$.
This lattice change approaching Cu-O-Cu angle of 90$^{\circ}$ realizes the shrinkage in the $b$ axis and the enhancement of the ferromagnetic exchange interaction in $b$ direction simultaneously, which is energetically favorable at the magnetized states.

This pantograph like lattice change is supported by theoretical estimation of exchange constant as a function of lattice changes based on DFT$+U$ calculations.
Besides, the magnetostriction dependence on magnetization curve with the relation of $\Delta L\propto M^{1.3}$ is roughly reproduced by calculations with Lancoz method in a finite lattice model, indicating that the strain in the quantum spin system is dependent on the spin-spin correlation within the exchange striction model.
Further discussion will be given elsewhere.

\subsection{2D quantum spin system on plaquett lattice: CaV$_{4}$O$_{9}$}
As discussed in the previous subsection, low dimensional quantum spin system attracts attention due to the variety of magnetic phases arising from the enhanced magnetic fluctuations due to the low dimensionality, quantum mechanical nature of spin.
Magnetic order is further suppressed leading to the enhanced magnetic fluctuation due to the frustrations arising from geometrical feature or competing exchange interactions.

The magnetic model of an antiferromagnet CaV$_{4}$O$_{9}$ is shown in Fig. \ref{cavo}(a).
It is a quantum spin system with spin-half on spin-lattice of interacting spin-plaquettes. 
Due to the strongest exchange interaction, spin-plaquettes forms.
The spin-gap is formed in each plaquette unit.
The ground state of the system is a gapped non-magnetic phase, which has a large gap of 100 K \cite{Taniguchi}.
With increasing external magnetic field up to megagauss fields, it is expected that the crossover from the spin singlet state to the magnetic spin-triplet and spin-quintet states are observed in each plaquette \cite{Fukumoto}.
It is further expected that some exotic phases such as magnetization plateau (magnetic solid), magnetic superfluid or magnetic supersolid phase may emerge, by further taking into account the inter-plaquette interactions and spontaneous symmetry breakings.

\begin{figure*}[t]
\centerline{\includegraphics[width=14cm]{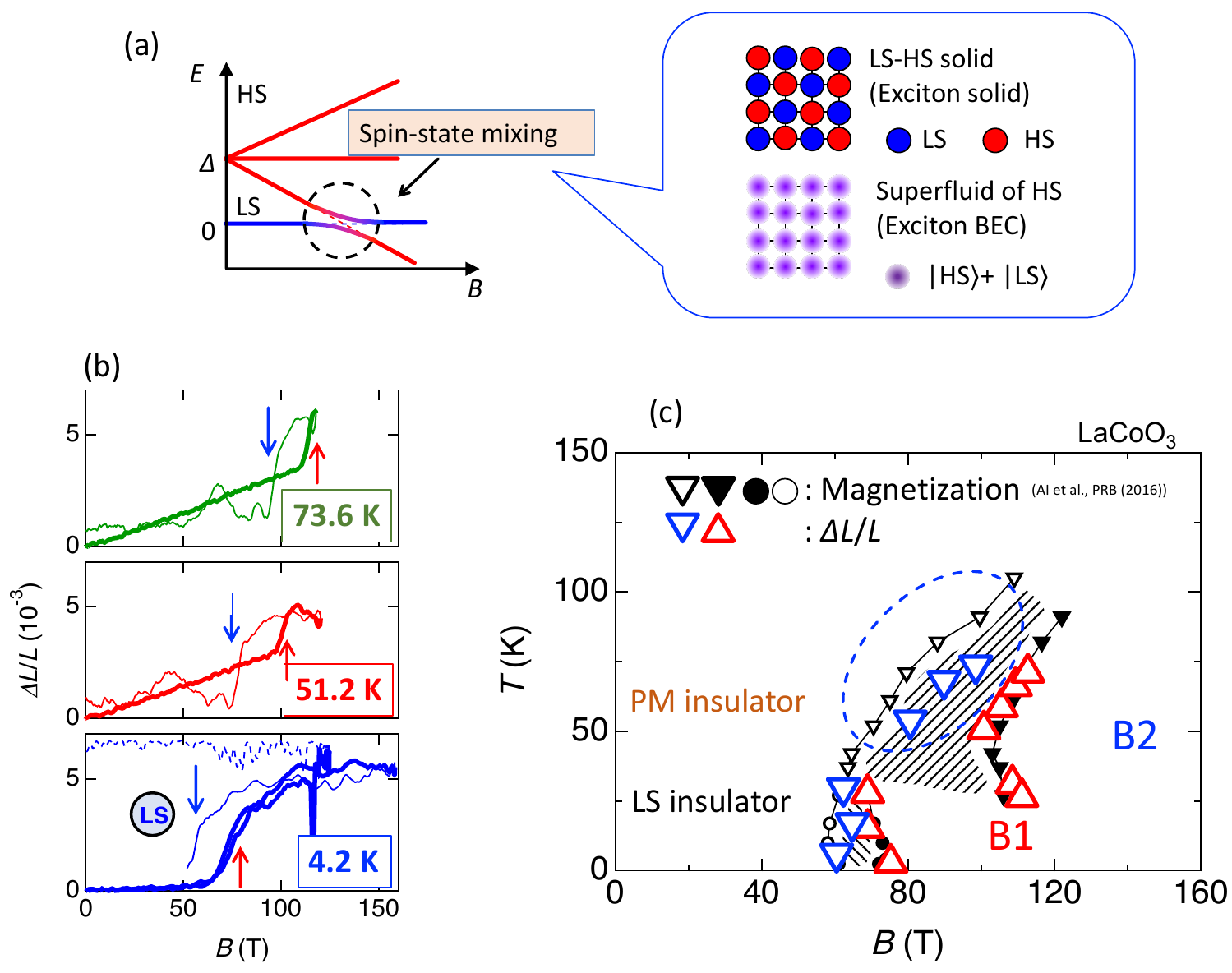}}
\caption{(a) Schematic drawing of the magnetic field effects on the spin-states and emerging exotic phases.
(b) Temperature dependence of magnetostriction curves up to megagauss fields.
(c) Magnetic field - temperature phase diagram of LaCoO$_{3}$ based on the previous magnetization measurement and the present magnetostriction measurements.
}
\label{lco}
\end{figure*}

We have carried out magnetization and magnetostriction measurements up to 150 T and 190 T, respectively, as shown in Figs. \ref{cavo}(b) and \ref{cavo}(c).
As show in the $dM/dB$ curve up to 150 T in Fig. \ref{cavo}(b) a collapse of the spin gap is evident by the sharp increase of $dM/dB$ starting from 85 T.
The $B_{\rm{C1}}=85$ T for the closing of the spin gap well coincides with a theoretical prediction \cite{Fukumoto} using magnetic parameters derived from the neutron scattering data \cite{Kodama}.
In further exploration using FBG with optical filter method, we find in the magnetostriction data a plateau region above 175 T.
Magnetostriction also starts to increase at 85 T indicating that the collapse of the spin-gap is re-confirmed.

As summarized in Fig. \ref{cavo}(d), the spin-gap state extends to 85 T at low temperatures below 10 K.
From 85 T to 175 T, a magnetic superfluid state may possibly be realized.
Above 175 T, a first magnetostriction plateau is found, which corresponds to the solidification of magnons.
In the exchange striction model, the magneotstriction is a function of magnetization.
Within the model, thus, the plateau in the magnetostriction corresponds to the magnetization plateau.
Note that the assignment of the magnetization plateau is ambiguous.
One reason for this is that the absolute value of the magnetization plateau is not concretely decided yet.
If the plateau corresponds to a half-magnetization plateau, it will be a trivial plateau with spin-triplet state for each plaquette.
If the plateau corresponds to a 1/4-magnetization plateau, it may be a non-trivial plateau with $\sqrt{2}\times\sqrt{2}$ superstructure of spin-triplet plaquettes on the sea of spin-singlet plaquettes, which emerges with a translational symmetry breaking.
Further magnetostriction process is a scope of future measurement with the EMFC method.

\subsection{Spin crossover oxide: LaCoO$_{3}$}
Due to the competing strength of Hund's coupling and crystal field splitting in cobalt oxides, many cobaltites undergo the spin-crossover where the spin-state of Co ion changes with the variation of external fields such as temperature, pressure, magnetic fields and transient photo-irradiations.
Among them, LaCoO$_{3}$ is by-far the most intensively studied material for its puzzling spin crossover process from the low-spin (LS) insulator ground state, then to the paramagnetic insulator state and then to the paramagnetic metallic state with increasing temperature.

Cobaltites can be viewed as a strongly correlated electron system with many electrons and multiple orbitals.
From the point of view, an interesting proposal has been claimed theoretically based on multi-orbital Hubbard models \cite{Kunes, Nasu}.
Let us consider the simplest case of the two-orbital and two electron system.
The spin-state change from the low-lying low-spin (LS) state with $S=0$ to the excited high-spin (HS) state with $S=1$ can also be viewed as a creation of an exciton with $S=1$ from the vacuum state within the field theoretical consideration.
The exciton has a nature of inter-site itineracy and the inter-exciton repulsive interactions.
As a result, at low temperatures, the system undergoes phase transitions to the solidification of excitons with translational symmetry breakings where the inter-exciton interactions are significant.
In the case where the intineracy of the excitons are significant, the superfluidity of excitons with $U$(1) gauge symmetry breaking may take place, which is called the excitonic condensation or excitonic insulator phase.

Such exotic phases may appear at high magnetic fields in LaCoO$_{3}$
At its ground state the low lying LS state is located.
The magnetic excited spin-state can be stabilized by the application of high magnetic fields as schematically drawn in Fig. \ref{lco}(a).
At around the $B_{\rm{C}}$ where the field induced spin crossover takes place, due to the spontaneous symmetry breaking, an inter spin-state hybridization may take place resulting in the emergence of excitonic condensation phase and HS-LS solid phase, where the former corresponds to the exciton superfluidity and the latter corresponds to the solidification of excitons.

We previously reported an experimental observation of unusual spin-state ordering at megagauss fields as shown on the $B$-$T$ phase diagram in Fig. \ref{lco}(c) using magnetization measurements \cite{IkedaPRB01}.
Later it is argued by several theoretical groups that the B1 and B2 phases correspond to the LS-HS solid phase and the excitonic condensation phase, respectively \cite{Tatsuno}.

For further investigation of the high field phases, we have carried out a series of magnetostriction measurement of LaCoO$_{3}$ up to megagauss fields with variation of the initial temperatures.
Representative results are shown in Fig. \ref{lco}(b).
A clear temperature dependence is observed where the transition magnetic field increases with increasing magnetic field.
The obtained transition fields are plotted using colored symbols on the previously constructed $B$-$T$ phase diagram as shown in Fig. \ref{lco}(c).
Overall, the transitions fields obtained in the magnetostriction study are well accordance with those of the magnetization study.
Note, however, a systematic deviation from the previous phase boundary is observed in the down sweep of magnetostriction at $T>30$ K.
Origin of this deviation is not clear at this moment.
Decoupling of magnetization and magnetostriction may indicate the decoupling of orbital state and spin state in cobalt ion.
Based on the spin-state picture, total spin and orbital state are very firmly coupled.
Such decoupling of lattice and magnetism is, thus, difficult to happen in the spin crossover system than the usual spin system.
Other possibilities are the extrinsic vibrations in the complex of the FBG and sample, and the non-equilibrium state in the sample realized in the down sweep of the pulse field.

In the present experiment, a single crystalline sample was used.
For future study, we are planning to perform experiments with a sintered power sample.
Sintered power samples are advantageous in the senses that the first order transition gets softer making the FBG measurement easier \cite{Sato2009}, and that one can avoid magnetic torque due to the anisotropic magnetization of the magnetic phase of LaCoO$_{3}$.

\section{Summary}
We have reviewed our recent development of the high-speed 100 MHz strain measurement system based on the FBG technique and the optical filter method devised for the magentostriction measurements at $\mu$s-pulsed magnetic fields for megagauss region.
Measurement system for the optical filter method is introduced.
As working examples, the measured results on solid oxygen, volborthite, CaV$_{4}$O$_{9}$ and LaCoO$_{3}$ are briefly reviewed.

\section*{Acknowledgment}

This work was supported by JSPS KAKENHI Grant-in-Aid for early career scientists Grants No. 16K17738 and No. 18K13493, Grant-in-Aid for Scientific Research (B) Grant No. 16H04009, and the internal research grant from ISSP, UTokyo.

\bibliography{mg16}
\bibliographystyle{IEEEtran}

\end{document}